\documentclass[conference]{IEEEtran}
\ifCLASSINFOpdf
\else
\fi
\usepackage{tikz}
\usepackage{amsmath}
\usepackage{mathrsfs}
\usepackage{graphicx}
\usepackage{enumerate}
\IEEEoverridecommandlockouts

\begin{document}
%
\title{Scaling Laws in Chennai Bus Network}

\author{\IEEEauthorblockN{Atanu Chatterjee and Gitakrishnan Ramadurai}
\IEEEauthorblockA{Transportation Engineering Division\\Department of Civil Engineering\\
Indian Institute of Technology\\
Madras, Chennai 600036\\}}


%


\maketitle

\begin{abstract}
In this paper, we study the structural properties of the complex bus network of Chennai. We formulate this extensive network structure by identifying each bus stop as a node, and a bus which stops at any two adjacent bus stops as an edge connecting the nodes. Rigorous statistical analysis of this data shows that the Chennai bus network displays small-world properties and a scale-free degree distribution with the power-law exponent, $\gamma = 3.8$.
\end{abstract}


%
\IEEEpeerreviewmaketitle

\section{Introduction}
\subsection{Chennai Bus Network}
Chennai is one of the metropolitan cities in India with a structured and a close-knit bus transport network. The Chennai bus network (CBN) is operated by the Metropolitan Transport Corporation (MTC), a state government undertaking. Spanning an area of 3,929 sq. km and with over 800 routes sprawling across entire Chennai, this extensive network also boasts of the largest bus terminus in Asia. With the population of the city being the sixth largest in the country, this medium of transport is most widely used for day-to-day commutation. \par
The reason that the bus network, in general, achieves this favourable status lies primarily in two factors: (i) $cost$, and (ii) $penetration$. With high volume of commuters flowing through this network each day, the cost of travel per day, per person becomes significantly low. Although, this is a parameter which is usually met by every other public transport system, the penetration factor is the one which significantly makes bus networks the prime mode of transportation in cities. A high penetration effect exhibited by a bus network enables a person to travel between any two points in the city very easily which is in contrast to other public modes of transportation, where geographical constraints associated with the network layout are much higher in number. Thus, bus transportation allows flexibility in travel. A natural question arises here; what is the least number of buses one should change to go from one place to another? Or more precisely, why do we require changing buses at all? To answer these questions, we need to look at a more abstract and networked structure of the bus routes, and argue based on its various statistical properties.
\subsection{Complex Networks: overview}
The recent surge of interest in the field of network science has caused researchers to formulate almost every physical system as an interacting network [1 - 8]. We define a network as a graph, $\mathscr{G}$, consisting of links, $L$, and nodes, $N$, in a tuple, $\mathscr{G} = (N, L)$ where, $N = (n_1, n_2,...)$ $\in$ $\Re$ and $L = N\times N$ $\in \Re^2$. When such a network shows non-trivial properties like the patterns in which the wiring between the nodes occur, then such a network is usually termed as a complex network [3 - 4]. The non-triviality in the properties arises primarily from the varying degrees of the strength of interaction between a pair of nodes, which is non-uniform and irregular. In the light of the success of recent investigations using a network science approach, we formulate the CBN as a complex network, and analyze its various statistical properties [9]. \par
The irregularity in the interaction patterns of nodes have been found in many social, biological and technological systems. Two prominent network models have been proposed in order to understand these systems: (a) \textit{small-world network model}, and (b) \textit{scale-free network model}. Both the models are categorized by certain network metrics, which are many, namely degree-distribution, clustering coefficient, network diameter, average path length, betweenness centrality, etc. The small-world network model was proposed by Watts and Strogatz in 1998 [10] and included a novel modification of the Erdos-Renyi random graph model which was proposed in 1959 [11]. The small-world network is an interesting model that captures our social interactions in a very promising way. The two major characteristics of this model are: (i) \textit{short average path length} between any pair of nodes, and (ii) \textit{high clustering effect} denoting strong-knit communities [8, 12]. \par
Small-world effects have been found in electric power grids [7], road maps [13], protein-yeast (metabolite) interaction networks [14], etc. Also, it is interesting to note that the Indian railway network (IRN) shows small-world characteristics [5]. The other model was proposed by Barabasi and Albert in 1999 when they mapped the topology of a portion of the World Wide Web [2]. An interesting aspect of their model is the \textit{scale-free degree distribution} of the nodes, marked by the presence of $hubs$ [1, 2]. Social networks [15], movie actors collaboration networks [16], authors citation networks [17], the World Wide Web [2], protein-protein interaction networks [18], mobile communication networks [19] etc. have been shown to follow a scale-free growth. With respect to transportation networks, airline networks and some subway networks have been observed to obey the Barabasi-Albert scale-free model [9, 20, 28, 29, 31]. However, it is worthwhile to note that the network topologies are not the absolute metrics that define strict classes of physical systems, e.g., some researchers have proposed a scale-free model for the rail-network of China [20].
\begin{figure*}[t]
\centering
\scalebox{0.4}
{\includegraphics{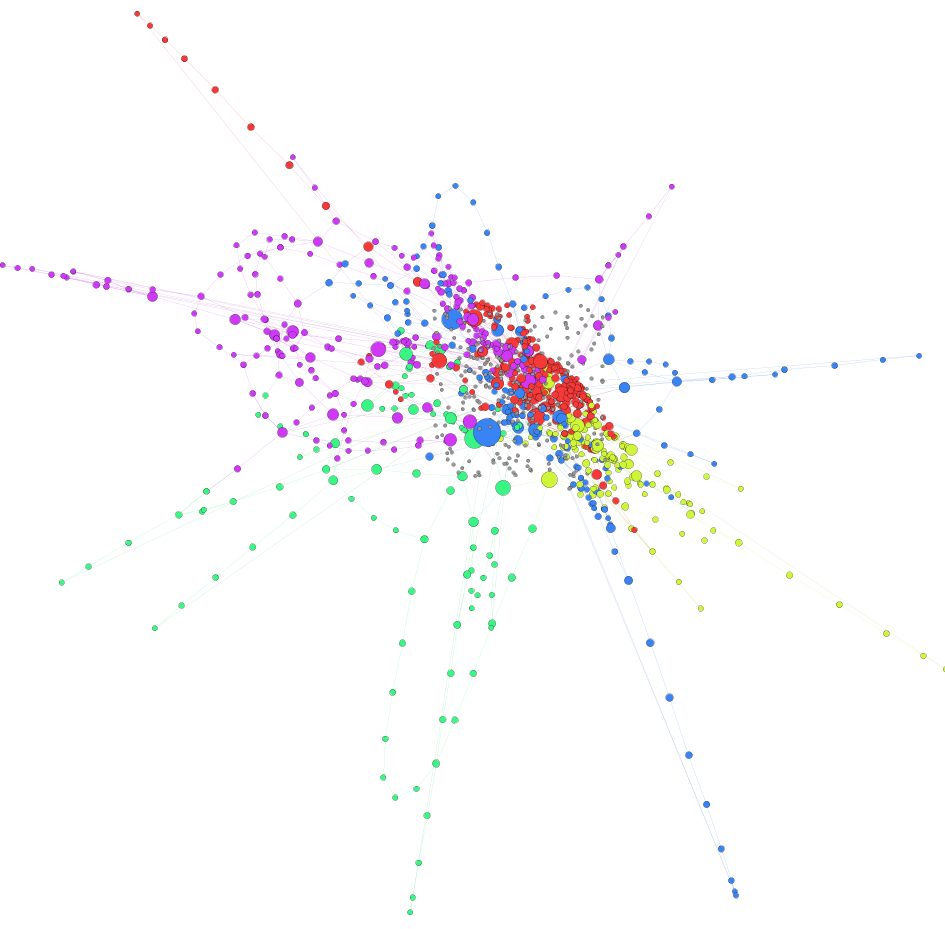}}
\caption{The Chennai Bus Network is plotted with nodes, $N = 1009$ and links, $L = 1610$. The layout is obtained using force-directed algorithm with complexity $\mathscr{O}(n^2)$. The node sizes vary according to the betweenness-centrality of the nodes. Color partioning is done in order to distinguish between the communities, as seen in figure, five communities have been identified.}
\end{figure*}
\section{Methodology}
Before creating an abstract model of the bus transport network, one needs to clearly distinguish between the geographical placing of the actual bus stops and the placing of the analogous nodes in a complex connected network. We define a connected network as a graph, in which a path traversal is possible between any pair of nodes. We also define the two analogous spaces: the \textit{geographical space} and its topological equivalent, \textit{network space}. In the network space, every node pair $(i-j)$ is separated by a distance, $l_{ij}$ of one unit. The actual significance of defining an equivalent network space lies in the value of this distance metric, $l_{ij}$. If $l_{ij} = 1$, there exists a bus that takes a commuter from $i$ to $j$ directly whereas, if $l_{ij} > 1$, then a person has to change buses at least once at some intermediate stop between $i$ and $j$. \par
The MTC website maintains a database of the entire bus route of Chennai. In order to create our network model, we identify every bus stop as a node and a bus connecting any two adjacent bus stops as a link between them. At this point, we assume that the bus network is bi-directional in nature hence, the links connecting a pair of nodes are undirected. In order to analyze this extensive network, a node-node adjacency matrix, $\mathscr{A}(N, L)$ is generated. An element of this matrix, $a_{ij}$ is either zero, one, or some quantity greater than one. If two nodes,$ i-j$ are not connected directly then, $a_{ij} = 0$. Since parallel edges are possible, $a_{ij}$ can take any integer positive value; note that $a_{ii} = 0$. The summation of the row, i.e., for all $i$ and a fixed $j$, we have the total degree of the $j^{th}$ node. The degree of a node implies the total number of links that pass through a node. For a node signifying a bus stop, as in this case, the degree of a bus stop will signify the strategic importance that bus stop manifests on the network. A bus stop with a high degree will have a high number of routes crossing it. Nodes with very high degrees in a network are often termed as $hubs$. Since our network is undirected, we do not distinguish between out-degree and in-degree for a node thus, $deg(j) = outdeg(j) + indeg(j)$. Further, we define a weighted graph in which the edges are assigned weights, $w$. In this case, weights are integer values denoting denoting the frequency of appearance of a particular link in the $800$ bus routes operating throughout Chennai. Thus, we define an analogous concept of weighted degree, $w-deg$ for every node, $i \in N$ as, the sum of all the link weights incident on that node, $w-deg(i) = \sum_{j}w_{ij} \forall$ $l_{i, j} \in L$. In order to analyze various properties of this complex network, we identify the metric such as the clustering coefficient, which is a measure of the degree to which the nodes have the tendency to cluster together. For an undirected network, the local clustering coefficient of node $i$ is given by, $C_{i} = 2\times l_{ij}/k_i\times (k_i - 1)$ where, $l_{ij}$ is a link connecting $i$ to $j$, and $k_i$ are the neighbours of node $i$. The neighbourhood, $N_i$, for a node, $i$ is defined the set of its immediately connected neighbours as, $N_i = \{ n_j : l_{ij} \in L \wedge l_{ji} \in L\}$. For the complete network, Watts and Strogatz defined a global clustering coefficient, $\mathscr{C} = \sum_i C_i/n$. The other important metric of particular interest is the degree-distribution function, $p(k_i)$, where $p(k_i)$ is the probability of degree, $i$, having a degree count, $k$. It is a measure of the degree distribution of nodes varying across the entire network. The shortest path between any pair of nodes is computed using \textit{Dijkstra's algorithm}.  
\begin{figure*}[t]
\centering
\scalebox{0.25}
{\includegraphics{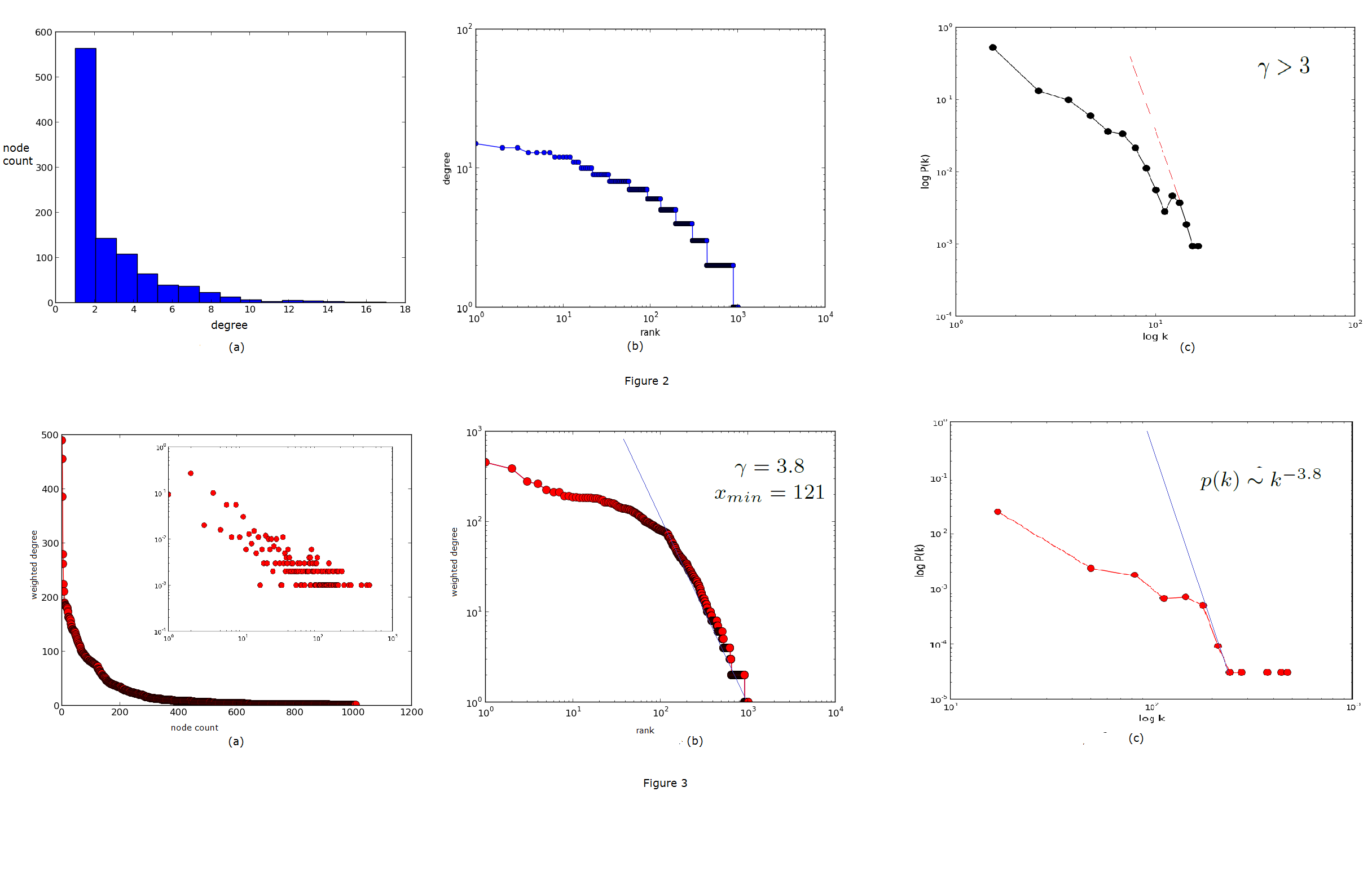}}
\caption{(a) A histogram plot showing the degree distribution and node count for an un-weighted network; (b) Rank-degree correlation plot on a log-log scale showing weak $Zipfian$ distribution; (c) Plot of \textit{log p(k)} vs. \textit{log k} for an un-weighted network, and the corresponding straight line fit for the tail with slope $\gamma > 3$}\caption{(a) Plot showing the weighted degree-distribution; weighted degree-distribution on a log-log scale ($inset$); (b) Weighted degree-rank plot on a log-log scale with $\gamma = 3.8$, and cut-off, $x_{min} = 121$; (c) Plot of \textit{log p(k)} vs. \textit{log k} for an the weighted network, and the corresponding straight line fit for the tail following the power-law, $p(k) \sim k^{-3.8}$}
\end{figure*}
\section{Results}
We construct the network from the MTC database with the number of nodes, $N =1009$ and the number of edges, $L = 1610$. The layout of the CBN in Figure 1 is obtained by using force-directed algorithm, assuming every node is connected by a spring. The prime idea is to minimize the energy of the system as every node tends to repel every other node (to which it is not connected directly) while being attracted by the elastic force of the spring-like edge. Using clustering algorithm, \textit{Chinese whispers} [21] we find that the CBN has close-knit clusters. Although few in number, they are located in those places of Chennai where mostly residential areas and business centres have conglomerated with time. Clusters represent a structure of the node set such that node traversal is possible from any one node to every other node in that set. Clusters in social networks represent high density ties between groups of people. In a transportation network, clusters represent the density of inter-dependency between the nodes, e.g., commercial areas like market places will tend to cluster with residential areas. But it is interesting to note that the clustering coefficient, $\mathscr{C} = 0.068$ for the CBN is significantly higher than a random network generated on the same edge set. When the clustering coefficient, $C(k)$ for a node, $k$ is plotted on a log-log scale (Figure 4(c)) a scale-free distribution is observed. \par
\begin{figure*}[t]
\centering
\scalebox{0.25}
{\includegraphics{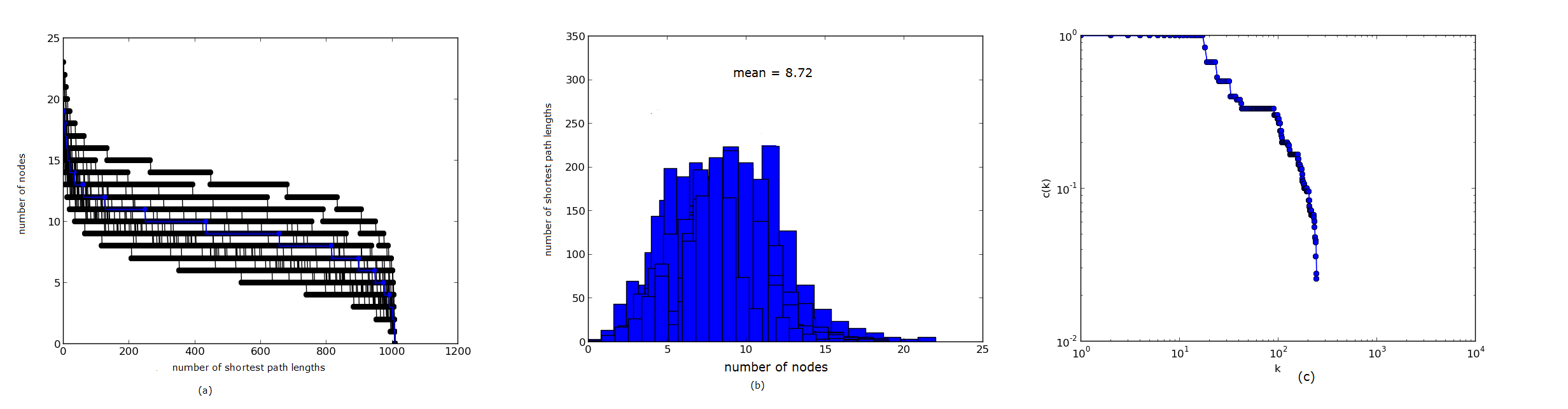}}
\caption{(a) Figure showing the variation of shortest path length for a single node to every other node with the number of nodes encountered; (b) Skewed $Gaussian$ distribution of the path length and number of nodes with mean, $\bar{l}_{ij} = 8.72$; (c) Figure represents the variation of clustering coefficient $C(k)$ with node count $k$}
\end{figure*}
Another important metric that quantifies the topological structure of a network is the degree distribution function, $p(k)$. Figure 2(a) represents the degree distribution of the CBN as a function of the number of nodes. It can be clearly seen that the histogram decays very rapidly initially, for all degrees, $d_i \leq 2$, but afterwards, the decay is uniform for degrees, $2 < d_i \leq 6$. Henceforth, the decay is again rapid for all degree, $d_i > 6$. It is also observed that the majority of the nodes have a very low degree while a very $few$ nodes have significantly high degrees. The average degree, $\bar{d}_i$ for the nodes in the CBN is found to be $3.19$ with the median degree, $\bar{D}_i$ being $2$. The maximum and minimum degrees for the CBN are $d_{max} = 15$ and $d_{min} = 1$. When the degree-rank correlation is plotted on a log-log scale, we get a weak $Zipfian$ distribution, as seen in Figure 2(b) [22]. From Figure 2(a) and 2(b), it is evident that the degree and rank of a node are inversely related. Therefore, the CBN shows evidence for a $power-law$ degree distribution, $p(k) \sim k^{-\gamma}$. In order to find the power law exponent, $\gamma$, the probability distribution function for a degree, $p(k_i)$, is plotted with respect to the degrees of nodes, $k_i$, on a log-log scale. Since it is presumed that the distribution follows a power-law curve, the $k-p(k)$ plot on a log-log scale should be a straight line. Figure 2(c) shows a linear trend in the scatter of points on the log-log plane. In order to view this more clearly we plot the weighted degree-distribution in Figure 3(a, b, c). It is clearly visible that the distribution of the points is clustered at the tail end of the line in the log-log plane. The slope of the straight line that captures the $fat$-tail of the degree distribution is the power-law exponent, $\gamma$. The slope of the straight line is found to be equal to $3.8$ [30]. \par
It is quite natural for a traveler to search for the shortest path from his point of origin to his destination. The shortest path between a pair of nodes is the one that minimizes the number of node traversals between them. In order to answer the above question, the shortest path from a single source to every other node is plotted in Figure 4(a).  The curve resembles a $S$-pattern. An interesting point to observe here is that more than $80 \%$ of the paths require $8$ or less nodes, and only about $20 \%$ of the remaining paths require more than $8$ nodes. This process is repeated for all the ${N}\choose {2}$ node pairs in the network, and the sigmoid distribution is found to be similar in all the cases. In Figure 4(b), we plot the normalized sigmoid curves. It essentially reveals a skewed Gaussian distribution of the shortest path lengths with the mean path length $\bar{l}_{ij} = 8.72$, which implies that on an average, any random pair of nodes can be visited by at most nine hops across this extensive bus network. Also, observe that the number of nodes steeply rises till $26$ for the remaining $20 \%$ nodes, accounting for the large diameter of the CBN. Also, it is worth noting that the average path length varies with number of nodes as $\bar{l}_{ij} \sim ln(N)$, which is a characteristic feature of a small-world network [23]. Finally, the heavy tail of the $C(k)-k$ is seen to follow the power law as, $C(k) \sim k^{-3.8}$. 

\section{Discussion}
The diffusion of structural and statistical concepts of network science has not been uniform in understanding transportation networks in general. While networks such as airlines, railways, and public transit networks have been extensively studied [6, 24-28 ], the quantum of work done in understanding road transportation, and in particular, the bus networks, has been relatively less [29]. In this paper, we have rigorously analyzed the topological structure of the bus network of Chennai. It is quite interesting to see that the CBN shows various remarkable characteristic properties.\par
The density of this extensive network is surprisingly low, $\rho \sim 0.003$, which gives us an idea of the spatial flow of the network across the entire city from the core to the extremities. As the CBN shows a scale-free degree distribution, there are a significantly large numbers of nodes that have a low degree when compared to the mean. On the contrary, majority of the links are held by a very few number of nodes, as we can see from Figures 2 $\&$ 3.  These set of nodes, which contribute towards majority of the connections in the network are often termed as hubs. Thus the CBN shows existence of hubs, which is again an interesting result for a transportation network. The core of the network, inclusive of the hubs, would often tend to cluster around the regions in the city which experience maximum flow of commuters. Also note that the existence of bus depots, where buses from all routes eventually converge to, can also be considered as hubs. \par
Another characteristic feature of this network that requires focus is the power-law exponent, $\gamma$. The value of $\gamma$ is found to be equal to $3.8$, with the cut-off, $x_{min} =121$. When $x < 121$ the degree-distribution follows an exponential growth while, $x > 121$ the degree-distribution curve shows a heavy-tailed growth. We also try to understand the dependence of $\gamma$ on the topological structure of the network, and correlate networks with similar characteristics.  A critical observation of the histogram plot in Figures 2(a) $\&$ 3(a) reveal before us the uniformity in the degree distribution towards the end of the tail. A high value of $\gamma$ represents more spread in the degrees amongst nodes, which is contrary to many real networks where $2< \gamma <3$, like the internet, power grids, social networks, etc., which have numerous hubs. In case of transportation networks, the value of $\gamma$ has been found to vary from $2$ to $5.5$ [28] for some metro networks, and $2< \gamma <3$ for public transit networks. A special thing to note here is that the networks for which $2< \gamma <3$, have been found to show ultra small-world characteristics, $\bar{l}_{ij} \sim ln((ln(N))$ whereas for $\gamma > 3$, the dependence of the average path length on the number of nodes follows as, $\bar{l}_{ij} \sim ln(N)$, resembling small-world phenomenon [23]. Some earlier works in bus networks of China show linear to exponential degree distribution [29]. For a public transport system, it is of immense importance that the travel times are reduced and people can efficiently commute from one place to another by changing minimum number of buses. \par
In the Indian context, people have shown that the Indian railway network is a small-world network [5]. However, they found that the degree distribution is not scale-free. With respect to airline networks, some earlier works have claimed that the Indian airline network follows a scale-free growth with $\gamma = 2.2$ [31]. All of these results claim universality in the structure of transportation networks. Being scale-free induces an additional advantage to the network by making it more robust and resilient to random failures [1-3, 7-8, 32]. It is indeed good to have scale-free, small-world transportation systems that would effectively reduce travelling time in cities, and preserve network connectivity in critical situations. 

\section{Conclusion}
Although various concepts of network science are being increasingly used as a tool for studying the underlying topology of networks, it still remains an elusive tool to visualize road transportation networks. In this paper, we have formulated a network model for the sixth largest city in India and understand its various structural attributes. The remarkable features of this extensive network include \textit{scale-free degree distribution}, \textit{short average path length} and \textit{high clustering coefficient}. It is being claimed that it would require at most nine hops to traverse any pair of random nodes in this network. Thus, the CBN exhibits small-world phenomenon and a scale-free growth with $\gamma = 3.8$. Based on the above attributes, it can be claimed that the CBN is essentially an efficient and a robust network.

\section*{Acknowledgment}
The author acknowledges the support from Center of Excellence in Urban Transport at Department of Civil Engineering, Indian Institute of Technology, Madras sponsored by Ministry of Urban Development, Government of India. The author also thanks Sushmitha Penta and Madhav Ch, for their help with data collection, and Priyatosh Mishra, the course TA for discussion and comments. Finally, the author is indebted to Prof. Balaraman Ravindran for motivating him to take up this project.



%

\begin{center}
\rule{5cm}{.5pt}
\end{center}

\noindent A. Chatterjee and G. Ramadurai, ``Scaling Laws in Chennai Bus Network," \textit{International Conference on Complex Systems and Applications} ICCSA 2014, France.\\

\noindent \textbf{Email}: atanu@smail.iitm.ac.in

\end{document}